\documentstyle[pre,aps,eqsecnum]{revtex}

\begin{document}                             

\title{
Elastic Scattering by Deterministic and Random Fractals: Self-Affinity of the
Diffraction Spectrum
}

\author{Daniel A. Hamburger-Lidar\cite{dani}}
\address{
Racah Institute of Physics and The Fritz Haber Center for Molecular Dynamics\\
The Hebrew University of Jerusalem, Jerusalem 91904, Israel\\
}

\maketitle

\begin{abstract}

The diffraction spectrum of coherent waves scattered from fractal supports is
calculated exactly. The fractals considered are of the class generated
iteratively by successive dilations and translations, and include
generalizations of the Cantor set and Sierpinski carpet as special cases. Also
randomized versions of these fractals are treated. The general result is that
the diffraction intensities obey a strict recursion relation, and become
self-affine in the limit of large iteration number, with a self-affinity
exponent related directly to the fractal dimension of the scattering
object. Applications include neutron scattering, x-rays, optical diffraction,
magnetic resonance imaging, electron diffraction, and He scattering, which all
display the same universal scaling.

\end{abstract}


\section{Introduction}

Scattering is one of the most important methods of observation of structural
properties of matter \cite{Born,Newton,Taylor,Child,Ziman,Teixeira}. Fractals,
on the other hand, have in recent years received enormous attention as models
for the structure of matter \cite{Mandelbrot,Avnir:book,Bunde,Stanley}. Thus
the relation between the two is of very general interest, as it provides an
essential connection between physical observables and the highly intriguing
fractal geometry of matter. The first to consider this relation was probably
Berry \cite{Berry:diffractals}, who calculated some important averages for
scattering by a random self-affine screen, coining the term ``diffractal'',
for waves that have encountered fractals. In subsequent works, it was
demonstrated that diffractals have properties that differ significantly from
``ordinary'' scattered waves. The central feature that distinguishes
diffractals from ordinary wave fields (where geometrical optics applies), is
that they are scattered by objects that have structure on all scales, in
particular near their wavelength $\lambda$. This fact gives rise to various
{\em scaling laws} of the diffraction spectrum, reflecting the fractal
structure of the scattering object. In contrast, in scattering from
crystalline material, characteristic interference (Bragg) {\em peaks} are
observed, which are related through their positions to the underlying crystal
structure.

In a large variety of fields scattering processes can be described by a {\em
Fourier transform} (FT), which relates the scattering amplitude to some local
density or potential.  Examples (to be dealt with in some detail in
Sec.\ref{methods}) include neutron and x-ray scattering \cite{Ziman}, optical
diffraction \cite{Born}, Nuclear Magnetic Resonance Imaging (MRI)
\cite{Callaghan}, electron scattering \cite{Taylor}, and Helium scattering
\cite{Benny:review1}. In all these cases the FT is an approximation, but its
generality and simplicity have rendered it by far the most widely used
approach to scattering problems. The FT is applicable if multiple scattering
and resonances can be neglected, which is typically the case under conditions
of high incidence energy. A large bulk of literature, theoretical as well as
experimental, exists on scattering in the FT approximation from {\em random}
fractals. The well-known result is that the intensity $I(q)$ decays as a
power-law of the momentum transfer $q$, with the exponent related to the
fractal dimension $D$ of the scatterer
\cite{Teixeira,Pfeifer:fractals,Kjems,Schmidt:1,Wong:1,Wong:2}.
Since they are naturally less abundant, much less attention has been devoted
to the scattering from fractals which can be constructed by a deterministic
set of iterative rules. Scattering from such fractals, as well as randomized
versions of them, will be the subject of this paper. The few examples include
Berry's \cite{Berry:diffractals} above-mentioned work; further, mainly in
optics, calculations on wave transmission \cite{Konotop,Sun} and Fraunhofer
diffraction \cite{Iemmi,Berger,Guojun,Uozumi}, on Cantor-bars, Koch fractals
or Sierpinski-carpet like media; in x-rays, numerical calculations on
scattering by a Menger sponge \cite{Schmidt:2}, and measurements on
diffraction from Cantor lattices \cite{Jarrendahl}. The most extensive
treatment is probably due to Allain and Cloitre
\cite{Allain:2,Allain:1,Allain:3}. In Ref.\cite{Allain:1}, these authors
reported an optical diffraction experiment on deterministically generated
Cantor bars and Vicsek fractals, and showed the resulting structure factor to
be self-similar. In Refs.\cite{Allain:2,Allain:3}, they analytically solved
and discussed properties such as band structure and scaling, for the
diffraction problem in the case of scattering from a class of fractals similar
to those to be discussed here. However, their discussion is essentially
limited to Fraunhofer diffraction and does not include random
fractals. Diffractal scattering for probes such as He-scattering or MRI
appears not to have been discussed in the literature. Thus, there seems to
have been no general treatment of the diffractal-FT problem.

The purpose of the present contribution is to demonstrate that an exact
solution for this problem is possible, in the case of scattering by objects on
an iteratively generated {\em fractal support} (see Fig.\ref{fig:2D-fractals}
for an illustration of the concept). The class of fractal objects that will be
considered here are those that can be generated by a combination of dilations
and translations.  Well-known examples of such objects include
(generalizations of) the Cantor set, Vicsek fractal, and Sierpinski carpet
\cite{frac-scatt:comment0}. An operator formalism will be introduced for this
purpose in Sec.\ref{generation}, which will allow the treatment of diverse
scattering conditions. It will appear in Secs.\ref{simple}-\ref{random} that
whereas some details of the diffraction spectrum are context-sensitive (i.e.,
determined by a form-factor), the overall structure is determined by a
universal, context-independent scaling relation. This conclusion is unaltered
(Sec.\ref{random}) by the introduction of a fractal dimension-preserving
randomness. Following this finding, Secs.\ref{form},\ref{SA-power} attempt to
address the connection between the results derived here for a somewhat
artificial class of fractals, and the standard (power-law) expressions used to
interpret scattering data from natural fractals, such as self-similarity
displaying porous materials, aggregates, or ramified structures.

The fractal scattering object can be generated in two essentially different
ways: from bottom-up (henceforth BU -- iterative inflation) or from top-down
(TD -- iterative refinement). In the former, the smallest unit remains fixed,
and structure appears at ever larger scales, limited of course by a natural
upper cut-off. This structure is reflected at ever {\em smaller} scales in
momentum space. A fixed point is reached where further spectral details are
indiscernible, either due to experimental resolution constraints or when the
wavelength becomes larger than the upper cut-off. To every iteration of the
fractal support there corresponds a diffraction spectrum. Subsequent
diffraction-spectrum iterations may be equated when the fixed point is
reached. The BU description is appropriate, e.g., in the case of a fractal
formed around a single nucleation center in a deposition process, as coverage
is increased. In the TD case, the total system size is fixed and structure
appears at progressively smaller scales, limited by a natural lower
cut-off. This description is probably more appropriate for the physical
formation of fractal structures by {\em removal} of material (pore-fractals
\cite{Pfeifer-Avnir:book}). If the wavelength $\lambda$ of the incident waves
is fixed, there will necessarily be another fixed point, where structure
develops below $\lambda$, and further fractal details are
indiscernible. Another possibility is that $\lambda$ becomes smaller than the
lower cut-off. In both cases subsequent iterations of the diffraction-spectrum
can then again be equated. Consequently, in both BU and TD cases, as will be
shown here, the diffraction spectrum becomes (approximately) {\em
self-affine}, and the self-affinity (or H\"{o}lder) exponent is simply related
to the fractal dimension of the scattering object. The central new result
derived here is that this conclusion is unaltered neither by the physical
identity of many scattering probes, nor by the introduction of a fractal
dimension-preserving randomness.

\section{Fourier-Transform Relations for Coherent Wave Scattering}
\label{methods}

The purpose of this section is to summarize the relation between the
structural properties of the scattering set and the observable diffraction
spectrum, for various physical examples to which the FT is applicable. The
ultimate goal is to show that in spite of the apparently very different way in
which the interaction (potential) between the wave and the scattering object
enters the formulation in each of the cases considered, there are certain
universal features in the scattering intensities, which reflect only the
underlying fractal geometry of the scatterer. The different examples are
presented below in increasing order of computational probe-object interaction
complexity. Thus, whereas neutron scattering (Sec.\ref{neutron}) involves
merely a discrete Fourier sum over the nuclear coordinates, electron
scattering (Sec.\ref{electron}) requires the FT of a potential which is a
functional of the local electron density, and He scattering (Sec. \ref{He})
necessitates the Fourier transformation of a functional of the interaction
potential itself. Yet, it should be emphasized that the results presented in
this work, all pertain exclusively to {\em local} interaction potentials (as
holds for all the examples considered below). Non-local potentials have been
successfully considered in the literature as well, mainly in low-energy
nuclear problems \cite{Watson,frac-scatt:comment4}.

Notation and conventions: The momentum transfer is denoted by ${\vec
q}$; the scattering amplitude by $f({\vec q})$; spatial vectors by ${\vec r} = (x,y,z)$. Elastic scattering
is assumed throughout. As emphasized in each of the subsections below, the FT
is essentially always the consequence of a high-energy approximation.

\subsection{Neutron Scattering}
\label{neutron}

Neutrons may couple by virtue of their spin to magnetic moments.  However, the
interaction of interest in the present context, i.e, which gives rise to a
Fourier integral, is with non-magnetic material, where neutrons are scattered
by the nuclei. Due to the extremely short range of the strong force, this
process is treated almost exactly in the Born approximation. The
neutron-nucleus interaction potential [see Eq. (\ref{eq:Born})] is essentially
a delta function (the ``Fermi pseudo-potential'' \cite{Ziman}), so that if the
nuclear positions are $\{ {\vec r}_i \} $, then

\begin{equation}
f({\vec q}) = C \sum_i e^{-i {\vec q} \cdot {\vec r}_i} .
\end{equation}

\noindent The accuracy of
this expression depends on the extent to which one may neglect incoherent
scattering due to isotopes, and inelastic diffraction due to variation of the
structure with time (thermal vibrations or atom diffusion).

\subsection{X-Rays}
The well known Laue derivation \cite{Kittel}, yields the relation

\begin{equation}
f({\vec q}) = \int d{\vec r} \: n({\vec r}) e^{i {\vec q} \cdot {\vec r}} 
\label{eq:X}
\end{equation}

\noindent between the local electron concentration $n({\vec r})$ and the x-ray
scattering amplitude. The
assumptions underlying the Laue derivation are essentially that the
polarization and electric field intensity are linearly and locally related by
the dielectric susceptibility $\chi({\vec r})$, which itself is
frequency-independent. Furthermore, at the inherently {\em high x-ray
frequencies}, $\chi \ll 1$, which allows for a decoupling of the equations
resulting from the attempt to solve the electromagnetic wave equation in the
crystal lattice, and yields Eq. (\ref{eq:X}).

\subsection{Optical Diffraction}
The FT arises in optics in the case of Fraunhofer diffraction. This holds when
both source and observation point are located very far from the aperture,
although some more general conditions exist \cite{Born}. The Fraunhofer
formula results from the {\em small-wavelength} Kirchhoff theory
\cite{Silver}, which solves the wave-equation under Huygens-Fresnel boundary
conditions. The assumed smallness of the optical wavelength in comparison with
the dimensions of the diffracting obstacles implies that in optical
diffraction, the BU fractal construction is more natural. Essentially,
Fraunhofer diffraction occurs when a coherent light wave is scattered by an
object with transmission function $t({\vec r})$, and the light amplitude is
obtained by a coherent superposition

\begin{equation}
f({\vec q}) = C \int d{\vec r} \: t({\vec r}) e^{i {\vec q} \cdot {\vec r}} .
\label{eq:optics}
\end{equation}

\subsection{Magnetic Resonance Imaging (MRI)}
Suppose the local nuclear spin density in a sample is $\rho({\vec r})$, and
that an oscillating magnetic field with local Larmor frequency $\omega({\vec
r})$ is applied to it. It is conventionally assumed in MRI that the Larmor
frequency is linear in the nuclear spin coordinates:

\begin{equation}
\omega({\vec r}) = \gamma |{\vec B_0}| + \gamma {\vec G} \cdot {\vec r},
\end{equation}

\noindent where $\gamma$ is the gyromagnetic ratio, and ${\vec B}_0$ is the
polarizing field, much larger than the linearly varying gradient field, of
which ${\vec G}$ is the gradient.  In practice, heterodyne mixing eliminates
the term $\gamma |{\vec B_0}|$, and the integrated MRI signal amplitude can be
written as

\begin{equation}
f(t) = \int d{\vec r} \: \rho({\vec r}) e^{i \gamma {\vec G} \cdot {\vec r} t}
.
\end{equation}

\noindent A reciprocal space vector ${\vec q} = \gamma {\vec G} t$ is
introduced \cite{Callaghan}, showing that ${\vec q}$-space may be
traversed by moving either in time or in gradient magnitude, so that

\begin{equation}
f({\vec q}) = \int d{\vec r} \: \rho({\vec r}) e^{i {\vec q} \cdot {\vec r}} .
\label{eq:MRI}
\end{equation}

\noindent Eq. (\ref{eq:MRI}) assumes {\em rapid signal acquisition} (after the
excitation pulse), so that spin relaxation, dipolar and scalar coupling, or
spin translation, do not distort the signal.

\subsection{Scattering of Electrons from Atoms}
\label{electron}

Here one often applies the Born approximation, 

\begin{equation}
f({\vec q}) = -{m \over 2\pi} \int d{\vec r} \: e^{-i {\vec q} \cdot {\vec r}}
V({\vec r}) ,
\label{eq:Born}
\end{equation}

\noindent valid at {\em high energies} and assuming that the electron (of mass
$m$) sees a fixed electrostatic potential due to a charge density $n({\vec
r})$,

\begin{equation}
V({\vec r}) = -e \int d{\vec r}' \: {n({\vec r}') \over |{\vec r}-{\vec r}'|} .
\label{eq:e-}
\end{equation}

\noindent This expression neglects the possible polarization of the
atom by the incident electron, as well as exchange effects \cite{Taylor}.

\subsection{He Scattering}
\label{He}

The He-surface scattering problem has been successfully treated within the
Sudden approximation \cite{Benny:Sud1,Benny:review2}, which assumes a {\em
high perpendicular momentum change compared to the momentum change parallel
to the surface} (essentially a high energy approximation). Under the presence
of an arbitrary He-surface potential $U({\vec R},z)$, the Sudden approximation
yields the scattering amplitudes as

\begin{equation}
f({\vec Q}) = {1 \over A} \int_A d{\vec R} \: e^{i {\vec R} \cdot {\vec
Q}}\,e^{2i \eta({\vec R})} ,
\label{eq:sud-general}
\end{equation}

\noindent where the phase-shift function is given in the WKB
approximation by

\begin{equation}
\eta({\vec R}) = \int_{\xi({\vec R})}^{\infty} dz\: \left[ \left( k_z^2 - {2m
\over \hbar^2} U({\vec R},z) \right)^{1/2} - k_z \right] - k_z \xi({\vec R}) . 
\label{eq:eta}
\end{equation}

\noindent Here ${\vec R} = (x,y)$, ${\vec Q} = (q_x,q_y)$, and $k_z$ is the
wavenumber component normal to the surface. The turning points $\xi({\vec R})$
are obtained as solutions to the energy equation

\begin{equation}
U[{\vec R},\xi({\vec R})] = {{\hbar^2 k_z^2} \over 2m} ,
\end{equation}

\noindent with $m$ the mass of the He atom. Effects such as resonances,
multiple collisions and dynamic polarization are neglected. For a hard-wall
potential

\begin{eqnarray}
U({\vec R},z) = \left\{ \begin{array}{ll}
	0        & \mbox{: $z \geq \xi({\vec R})$} \\
	{\infty} & \mbox{: $z < \xi({\vec R})$} ,
	\end{array}
\right. \nonumber
\end{eqnarray}

\noindent so that from Eq. (\ref{eq:eta}) it follows that in this case:

\begin{equation}
\eta({\vec R}) = -k_z \, \xi({\vec R}),
\label{eq:hw}
\end{equation}

\noindent as in the eikonal approximation in optics.

\section{Generation of Functions on Fractal Sets by Dilation and
Translation Operators}
\label{generation}

Having seen the generality of the FT in diffraction problems, the generation
of the scattering fractal support is given next. The construction to be
described below is in the spirit of the Iterated Function System formalism of
Barnsley \cite{Barnsley}.

\subsection{Simple Example}

Consider first as an introductory example the construction of a characteristic
function on the usual (ternary) Cantor set (Fig.\ref{fig:Cantor-step}, left):
One first contracts the generator (zero-order iteration), $\xi_0(x)=l$
($0<x<L$), by a factor $3$, and then places one copy of the contracted version
at the origin, and another translated by $2L/3$ from the origin. This
can be generalized to contractions by a factor $1/s$ ($0<s<1$) and
translations by $a\,L$. The
corresponding TD {\em fractal operator} is (the reason for using the
adjoint will become clear in Sec.\ref{simple}):

\begin{equation}
{\cal F}^{\dag} = (\openone + {\cal T}_{-a})\, {\cal C}_{1/s},
\label{eq:Fd}
\end{equation}

\noindent where the {\em translation operator} is defined as

\begin{equation}
{\cal T}_{a} h(x) = h(x+a L) ,
\label{eq:T}
\end{equation}

\noindent and the {\em dilation operator} is defined as

\begin{equation}
{\cal C}_{s} h(x) = h(s\,x).
\label{eq:C}
\end{equation}

\noindent ${\cal T}_{a}$ shifts the function it operates on by an
amount $a\,L$ to the left, and ${\cal C}_{s}$ stretches the function
by a factor of $1/s$. When applied in the inverse sense as required by
the definition of ${\cal F}^{\dag}$, it is easily checked that
$\xi_1(x) \equiv {\cal F}^{\dag}
\, \xi_0(x) = \xi_0(x/s) + \xi_0[(x-a L)/s]$, and that $\xi_n(x) = ({\cal
F}^{\dag})^n \, \xi_0(x)$ is indeed an $n^{\rm th}$ iteration stepped
Cantor surface, as illustrated in Fig.\ref{fig:Cantor-step}. Barnsley
\cite{Barnsley} and Vicsek \cite{Vicsek} provide a general theorem for
the calculation of the fractal dimension $D$ of such iteratively
constructed fractals; $D$ is the solution of the equation

\begin{equation}
\sum_{i=1}^n s_i^D = 1
\label{eq:D}
\end{equation}

\noindent where $s_i$ are all the contraction factors. Thus in the present
case:

\begin{equation}
\sum_{i=1}^2 s^D = 1 \:\: \Longrightarrow \:\: D = {\ln(2) \over \ln(1/s)}
\label{eq:D-here}
\end{equation}

\noindent To derive the algebraic
properties of the above operators, it is convenient to express them in
exponential form. ${\cal T}_a$ has the well known momentum-operator
representation

\begin{equation}
{\cal T}_{a} = e^{a L \, \partial_x}.
\end{equation}

\noindent This can be used to find a similar representation for ${\cal C}_s$:
Let $\mu = \ln(s)$, $y = \ln(x)$, and $g(y) =
h(x)$. The argument of $h(s\,x)$ can then be expressed in terms of a
sum: $h(s\,x) = h[\exp(y+\mu)] = g(y+\mu)$. But this is exactly in the form of
a translation, so that using the representation of ${\cal T}_a$ one finds:
$g(y+\mu) = \exp(\mu\,\partial_y) g(y)$. Noting that $\partial_y =
\partial_{\ln(x)} = x\,\partial_x$, one obtains the desired representation:

\begin{equation}
{\cal C}_{s} = e^{\ln(s) \, x\,\partial_x}.
\end{equation}

\noindent From here, using $\partial_x^{\dag} =
-\partial_x$ and $\partial_x x =
1+x\,\partial_x$, it is easily seen that

\begin{eqnarray}
{\cal T}_{a}^{\dag} = {\cal T}_{-a} \nonumber \\
{\cal C}_{s}^{\dag} = {1 \over s}{\cal C}_{1/s} .
\label{eq:daggers}
\end{eqnarray}

\subsection{General Construction of Functions on Fractals}
\label{general-construction}

The above formalism for {\em TD} fractals can easily be extended
to arbitrary dimension, as well as to {\em BU} fractals  Let
${\vec r} = (x_1,..,x_d)$ be a vector in $d$ dimensions. Then the
generalization of the 1D translation and dilation operators is

\begin{eqnarray}
{\cal T}_{\vec a} h({\vec r}) = h({\vec r}+{\vec a}L) \nonumber \\
{\cal C}_{s} h({\vec r}) = h(s\,{\vec r}).
\label{eq:TC-nD}
\end{eqnarray}

\noindent In exponential representation, it is easily seen that:

\begin{eqnarray}
{\cal T}_{\vec a} = e^{L {\vec a}\cdot\nabla} \nonumber \\
{\cal C}_{s} = e^{\ln(s) \, {\vec r}\cdot\nabla} .
\end{eqnarray}

\noindent A very wide class of fractals can be generated by using a
single contraction factor $s$ \cite{frac-scatt:comment5}: 

\begin{equation}
{\cal F}^{\dag} = (\openone + \sum_{i=1}^k {\cal T}_{-{\vec a}_i})\, {\cal
C}_{1/s} .
\label{eq:Fd-nD}
\end{equation}

\noindent For example, the Vicsek fractal \cite{Vicsek}, results by choosing\\
$s=1/3; \: \{{\vec a}_i\} = \{(2/3,0),(1/3,1/3),(0,2/3),(2/3,2/3)\}$,\\
whereas the Sierpinski carpet is generated by\\
$s=1/3; \: \{{\vec a}_i\} =
\{(1/3,0),(2/3,0),(0,1/3),(0,2/3),(1/3,2/3),(2/3,1/3),(2/3,2/3)\}$
(Fig.\ref{fig:2D-fractals}). Eq. (\ref{eq:D}) for the calculation of the
fractal dimension applies again, and one obtains in the present case:

\begin{equation}
\sum_{i=1}^{k+1} s^D = 1 \:\: \Longrightarrow \:\: D = {\ln(k+1) \over
\ln(1/s)}
\label{eq:D-here-nD}
\end{equation}

The BU fractal is most easily derived by employing the general fractal
operator [Eq. (\ref{eq:Fd-nD})], and the observation that repeatedly {\em
expanding} the TD fractal achieves the desired result. Thus the
general BU fractal-operator is:

\begin{equation}
{\cal G}^{\dag}_n = ({\cal C}_s)^n \, ({\cal F}^{\dag})^n .
\label{eq:Gd}
\end{equation}

\noindent Note that with this definition, it is guaranteed that the
smallest building-block making up the fractal is of unit length. Since
the expansion is one-sided, the fractal thus obtained is semi-infinite.

For future reference it is convenient to note, using Eqs.(\ref{eq:daggers})
for ${\cal T}_{\vec a}^{\dagger}$ and ${\cal C}_{s}$, that:

\begin{eqnarray}
{\cal F} = s^d {\cal C}_s (\openone + \sum_{i=1}^k {\cal T}_{{\vec a}_i})
\nonumber \\
{\cal G}_n = s^{-d\,n} {\cal F}^n {\cal C}_{1/s}^n .
\label{eq:F-G}
\end{eqnarray}

\section{Introductory Example: 1D, Hard-Wall H\lowercase{e} Scattering from a
Cantor Set}
\label{simple}

With the fractal operators defined, a simple, but prototypical diffractal-FT
problem can now be discussed. One may, e.g., consider 1D He scattering in the
presence of a hard-wall potential [Eqs.(\ref{eq:sud-general}),(\ref{eq:hw})],
with the shape-function

\begin{eqnarray}
\xi_n(x) = \left\{ \begin{array}{ll}
	l        & \mbox{: $x \in C_n$} \\
	0	 & \mbox{: else} .
	\end{array}
\right. \nonumber
\end{eqnarray}

\noindent $C_n$ denotes the $n^{\rm th}$ approximation to the Cantor set.

\subsection{Calculation of the Intensity Distribution}

\subsubsection{TD Case}

Denoting the phase-shift of a He atom with perpendicular wavenumber $k_z$ and
striking a step of height $h$, by

\begin{equation}
\Phi = -2k_z\,l ,
\end{equation}

\noindent one notes that $\exp(i \Phi \xi_n(x)/l) = \exp(i \Phi)$ for
$x \! \in \! C_n$ and $1$ otherwise. This calls for a normalized
characteristic function on the Cantor set. Such a function is just ${1
\over l} ({\cal F}^{\dag})^n\,\xi_0(x)$. Therefore the scattering amplitude is
(Eq. (\ref{eq:sud-general}) in 1D),

\begin{eqnarray}
f_n(q) = {1 \over L} \int_0^L dx\: e^{i q x} {e^{i \Phi} \over l} [({\cal
F}^{\dag})^n \xi_0(x)] + {1 \over L} \int_0^L dx\: e^{i q x} (1-{1 \over l}
[({\cal F}^{\dag})^n \xi_0(x)] ) =
\nonumber \\
{{e^{i \Phi}-1} \over {L\,l}} \int_0^L dx\: [{\cal F}^n e^{i q x}]
\xi_0(x) + {1 \over L} \int_0^L dx\: e^{i q x}  \xi_0(x) .
\label{eq:S-inter}
\end{eqnarray}

\noindent The last term is evidently just the specular contribution, and will
henceforth be assumed subtracted out. The penultimate term contains the
fractal operator, which in the present case equals [Eq. (\ref{eq:F-G})]

\begin{equation}
{\cal F} =  s {\cal C}_s (\openone + {\cal T}_{a}).
\end{equation}

\noindent What remains is to calculate ${\cal F}^n e^{i q x}$:

\begin{eqnarray}
{\cal F} e^{i\,q\,x} = s {\cal C}_s \left[ \left(1+e^{i q\,a L}\right) e^{i
q\,x} \right] = s \left( 1+e^{i q\,a L} \right) e^{i q\,s\,x} \nonumber \\
{\cal F}^2 e^{i\,q\,x} = s \left( 1+e^{i q\,a L} \right) {\cal F} e^{i
q\,s\,x} = s^2 \left( 1+e^{i q\,a L} \right) \left( 1+e^{i q\,s\,a L} \right)
e^{i q\,s^2\,x} ,
\label{eq:apply-F}
\end{eqnarray}

\noindent from which the general pattern can be inferred:

\begin{equation}
{\cal F}^n e^{i\,q\,x} = s^n e^{i q\,s^n\,x} \prod_{j=1}^n \left( 1+e^{i q\,a
L\,s^{j-1}} \right) .
\label{eq:F^n}
\end{equation}

\noindent This prototypical expression, or slight variants of it, will appear
repeatedly when more complicated cases are treated in later sections.
Before the intensities are obtained, the question of normalization
must be addressed. Since the Cantor set and its generalizations
discussed here have measure zero, the intensity is expected to vanish.
This can be avoided if the {\em intensity} is normalized to the relative
length occupied by the Cantor set support at the $n^{\rm th}$
iteration. There are $2^n$ steps in the set, each of length $s^n L$,
resulting in a normalization factor of $L(2 s)^n/L$.

The integration leading to the scattering amplitude [Eq. (\ref{eq:S-inter})]
can now be performed, yielding, after normalization:

\begin{equation}
f_n(q) = {1 \over {(2s)^{n/2}}} { {e^{i \Phi}-1} \over {i q\,L} } \left(e^{i
q\,s^n\,L}-1\right) \prod_{j=0}^{n-1} \left( 1+e^{i q\,a L\,s^j} \right) .
\label{eq:S-final}
\end{equation}

\noindent The last result bears some resemblance to the (complex-)
Weierstrass-Mandelbrot function \cite{Mandelbrot},

\begin{equation}
W(q) = (1-w^2)^{-1/2} \sum_{j=-\infty}^{\infty} w^j \left( e^{2\pi
i\,s^j\,q}-1 \right),
\end{equation}

\noindent which suggests that the off-specular amplitude, as well as the
intensity,

\begin{equation}
I_n(q) = |f_n(q)|^2 ,
\label{eq:I}
\end{equation}

\noindent may be {\em self-affine} functions. Before this is investigated,
consider first the BU construction.

\subsubsection{BU Case}

Essentially, all that needs to be done is to replace the TD operator
${\cal F}^n$ in the previous subsection, everywhere by the BU operator
${\cal G}_n$. From Eq. (\ref{eq:F-G}) this operator is in the 1D case:

\begin{equation}
{\cal G}_n =  s^{-n} {\cal F}^n {\cal C}_{1/s}^n .
\end{equation}

\noindent When this is applied to the Fourier basis-set one finds:

\begin{equation}
{\cal G}_n e^{i q x} = s^{-n} {\cal F}^n e^{i s^{-n} q x} = \left[
\prod_{j=1}^{n} \left( 1+e^{i s^{j-n-1} q\,a L } \right) \right] e^{i q x} ,
\label{eq:F^n-BU}
\end{equation}

\noindent where the last equality follows from the general result for
${\cal F}^n e^{i q x}$ [Eq. (\ref{eq:F^n})]. As for normalization,
since the fractal grows indefinitely in the BU case, it is most
convenient to normalize the intensity by the number of elementary
units. This is $2^n$ for the $n^{\rm th}$ iteration.

In anticipation of the more general treatment of Sec.\ref{general}, the
scattering amplitude $f_n(q) = {1 \over L} \int_0^L dx\: \exp(i q\,x)
\exp[-2i\,k_z\,\xi_n(x)]$ can now be written as

\begin{eqnarray}
f_n(q) = \left[ {1 \over 2^{n/2}} \prod_{j=0}^{n-1} \left( 1+e^{i s^{j-n} q\,a
L } \right) \right] F(q) \nonumber \\
F(q) = {1 \over L} \int_0^L dx \: e^{i q x} \phi_0(x) \nonumber \\
\phi_0(x) \equiv e^{-2i k_z\,\xi_0(x)} ,
\label{eq:S-final-BU}
\end{eqnarray}

\noindent where $F(q)$ can be interpreted as a {\em form factor} and
the term in square brackets as a {\em structure
factor} $S(q)$ \cite{frac-scatt:comment1}.

\subsection{Recursion Relation and Self-Affinity of the Off-Specular
Intensity Distribution}
\label{S-A}

Similarly to the fractal sets described above, self-affine functions can be
constructed iteratively, for example as deterministic models of random walks
\cite{Pfeifer-Avnir:book,Vicsek,Stanley:2}.  At each stage, a function of this
type satisfies the recursive scaling relation

\begin{equation}
h_{n+1}(x) = b^{-\alpha} h_n(b\,x) ,
\label{eq:Holder}
\end{equation}

\noindent and becomes rigorously self-affine in the limit $n \!\rightarrow\!
\infty$. $\alpha$ is denoted the H\"{o}lder, or self-affine exponent
\cite{Stanley:2}. An analogous recursion relation will now be derived for the
off-specular amplitudes and intensities $I_n(q)$ of the previous subsections.
In the $n \!\rightarrow\! \infty$ limit, these are therefore also self-affine
functions.

\subsubsection{TD Case}

Using the result derived previously for the scattering amplitude
[Eq. (\ref{eq:S-final})], the intensity satisfies

\begin{equation}
I_n(q) = {1 \over {(2s)^n}} \left({2 \over {q\,L}}\right)^2 [1-\cos(\Phi)]
[1-\cos\left(q\,s^n\,L\right)] \left[ 2^n \prod_{j=0}^{n-1} \left[
1+\cos\left(q\,a L\,s^j\right) \right] \right].
\label{eq:I_n(q)-TD}
\end{equation}

\noindent The recursion-scaling relation follows once it is recognized that
the scale factor $b$ from Eq. (\ref{eq:Holder}) is the dilation factor $s$ in
the present case:

\begin{equation}
I_{n+1}(q) = s\,I_n(s\,q) [1+\cos(q\,a L)] .
\label{eq:scaling-1}
\end{equation}

\noindent Clearly, due to the presence of the cosine factor, this is not in the
form of the self-affine scaling relation of Eq. (\ref{eq:Holder}), where a {\em
constant} factor multiplies the $n^{\rm th}$ iteration. However, in the TD
case, successive fractal iterations will result in successive diffraction
spectra that differ at ever larger $q$ scales. $q_{max}$, the largest possible
$q$, is fixed by energy conservation, irrespective of the structure of the
scattering fractal set. Therefore, when the finest fractal detail, $\Delta
x_n$, becomes smaller than $2\pi/q_{max}$, it becomes physically reasonable to
equate successive iterations. For these to match in the sense of
Eq. (\ref{eq:Holder}), the simplest criterion is to require equality of the
intensities in the vicinity of the specular, $q \rightarrow 0$ (at the price
of mismatch increasing with $q$). Proceeding thus, Eq. (\ref{eq:scaling-1})
will be in the form of the self-affinity relation [Eq. (\ref{eq:Holder})] if
$\cos(q\,a L)$ is evaluated at $q=0$. For then one finds

\begin{equation}
I_{n+1}(q) \approx s^{-\alpha_1}\,I_n(s\,q)
\label{eq:scaling-2}
\end{equation}

\noindent where

\begin{equation}
\alpha_1 = D-1
\label{eq:alpha}
\end{equation}

\noindent {\em with $D$ the fractal dimension of the Cantor set},
Eq. (\ref{eq:D-here}). Thus, the self-affinity exponent of the intensity
spectrum is related to the fractal dimension of the object scattered from. The
reason for the specific form of the expression for $\alpha_1$ will become clear
in Sec.\ref{scaling}. The accuracy with which Eq. (\ref{eq:scaling-2})
produces the required scaling can be seen in Fig.\ref{fig:self-affine}.
Plotted there are the intensities for He scattering from a 1D,
hard-wall step function arrangement on two different Cantor set supports (see
caption for details). Significantly, the intensities of {\em all maxima} (not
just the specular, corresponding to $q=0$) are accurately reproduced. This
situation can only be expected to improve as $n$ is increased, demonstrating
the self-affinity of the spectrum.

\subsubsection{BU Case}

The scaling relation in this case is somewhat different from the
TD fractal. From the scattering amplitude calculated for the BU
fractals [Eqs.(\ref{eq:S-final-BU})] one finds

\begin{equation}
{I_{n+1}(q) \over I_0(q)} = [1+\cos(s^{-1} q\,a L) ] {I_n(q/s)
\over I_0(q/s)} ,
\label{eq:scaling-1-BU}
\end{equation}

\noindent implying that scaling is obeyed to within the form-factor
(i.e, only the structure-factor, not the intensity, is fully
scale-invariant). For a BU fractal, features in successive diffraction
spectra develop at ever {\em smaller} $q$ scales. Beyond the
experimental $q$-space resolution, it is physically reasonable, as in
the TD case, to compare successive iterations, and to require the
intensities in the vicinity of the specular ($q \rightarrow 0$) to be
equal. Substituting $1$ for $\cos(q\,a L)$, it is now found that

\begin{equation}
{I_{n+1}(q) \over I_0(q)} \approx s^{-\alpha_2}\,{I_n(q/s) \over I_0(q/s)}
\label{eq:scaling-2-BU}
\end{equation}

\noindent where

\begin{equation}
\alpha_2 = D
\label{eq:alpha-BU}
\end{equation}

\noindent with $D$ again the fractal dimension of the Cantor set,
Eq. (\ref{eq:D-here}). This is demonstrated in
Fig.\ref{fig:self-affine:BU}, where the scaling recipe with $\alpha_2$ is seen
to hold with high accuracy.

\subsubsection{Numerical Check of the Self-Affinity}

To further test the self-affinity, the H\"{o}lder exponents of the structure
factors for He scattering from hard-wall step functions on an $n=8$ ternary
and $n=6, s=1/5$, $a=4/7$ Cantor set (Fig.\ref{fig:self-affine:BU}) were
calculated, using the reliable and accurate {\em epsilon-variation}
method \cite{Dubuc,comment-Dubuc}. The result is shown in
Fig.\ref{fig:variation}. The respective regression-slopes of 1.367 and 1.59,
from which self-affinity exponents of $2-1.367=.633$ and $2-1.59=.41$ are
obtained, compare favorably with the prediction of Eq. (\ref{eq:alpha-BU}),
yielding $\alpha_2=\ln(2)/\ln(3)=0.631$ and
$\alpha_2=\ln(2)/\ln(5)=0.43$. Significantly, the log-log plots are straight
over two orders of magnitude, and the higher order iteration ($n=8$) yields a
more accurate exponent. Note further that this experimental-like analysis
yields the same self-affinity exponent as the scaling analysis leading to
Eq. (\ref{eq:alpha-BU}), {\em without the $q=0$ approximation}.

In conclusion of this section, the analysis of both TD and BU
fractals suggests that {\em (1) the scattering intensity from a fractal
surface is itself (approximately) self-affine, and (2) the fractal dimension
of the scattering surface manifests itself simply through the H\"{o}lder
exponent of the scattering intensity}.  Hence an analysis of the scaling
properties of the scattering intensity should reveal if the scattering surface
is fractal, and if so, what its fractal dimension is.

In the following sections it will be shown that this conclusion holds for the
general (arbitrary local potential, any dimension) diffractal-FT
problem. However, first a commutation property of the operators under
discussion must be established. This property will make it possible to
demonstrate that the scaling discussed above is indeed independent of the
nature of the scattering probe, and is instead exclusively determined by the
geometry of the fractal scattering object.

\section{Commutation Property of Change-of-Variable Operators}
\label{COVO}

The translation and dilation operators ${\cal T}_a$ and ${\cal C}_s$ can
clearly be regarded from their definition [Eqs.(\ref{eq:T}),(\ref{eq:C})], as
``{\em change-of-variable operators}'' (COVO). Let ${\cal P}$ be a general
COVO, i.e.,

\begin{equation}
{\cal P} p = p' \: : \:\:\: p'(x) = p[\phi(x)]
\label{eq:COVO}
\end{equation}

\noindent The purpose of this short technical section is to prove that the
following commutation relation holds for COVO:

\begin{equation}
{\cal P}\{f[g(x)]\} = f[ ({\cal P} g)(x) ]
\label{eq:commutation}
\end{equation}

\noindent To prove this, consider the LHS: Let

\begin{equation}
f[g(x)] = h_0(x),
\label{eq:fg}
\end{equation}

\noindent and note that the LHS is ${\cal P} h_0 = h_0'$. But, by
Eqs.(\ref{eq:COVO}),(\ref{eq:fg}),

\begin{equation}
h_0'(x) = h_0[\phi(x)] = f\{g[\phi(x)]\} = f[g'(x)] = f[ ({\cal P} g)(x) ] ,
\label{eq:h1}
\end{equation}

\noindent which is identical to the RHS of Eq. (\ref{eq:commutation}), so that
the commutation property holds.

Note also that nothing in the above discussion restricted the result
to 1D: the commutation property holds in arbitrary dimension. Thus a
useful corollary follows immediately. Let $\partial G(x,z)/\partial z
= g(x,z)$. Then by the commutation property:

\begin{eqnarray}
{\cal P} \int_{\xi(x)}^{\zeta(x)} g(x,z) dz = {\cal P} \{ G[x,\zeta(x)] -
G[x,\xi(x)]\} = G\{\phi(x),\zeta[\phi(x)]\} - G\{\phi(x),\xi[\phi(x)]\} ;
\nonumber \\
\int_{{\cal P} \xi(x)}^{{\cal P} \zeta(x)} {\cal P} g(x,z) dz =
\int_{\xi[\phi(x)]}^{\zeta[\phi(x)]} g(\phi(x),z) dz =
G\{\phi(x),\zeta[\phi(x)]\} - G\{\phi(x),\xi[\phi(x)]\} \nonumber ,
\end{eqnarray}

\noindent so that:

\begin{equation}
{\cal P} \int_{\xi(x)}^{\zeta(x)} g(x,z) dz = \int_{{\cal P} \xi(x)}^{{\cal P}
\zeta(x)} {\cal P} g(x,z) dz .
\label{eq:P-int}
\end{equation}

\section{General Deterministic Diffractal-Fourier Transform Problem}
\label{general}

The tools are now prepared to consider the general, deterministic,
diffractal-FT problem. This will require the use of the commutation property
of change-of-variable operators to treat the variety of scattering probes
discussed in Sec.\ref{methods} in a unified way.

\subsection{Structure of the Fourier Integral}

The ingredients entering the general problem are (1) the FT relations
from Sec.\ref{methods}, and (2) the realization that in every such
instance, the fractal structure may be introduced into the problem by
the repeated application of fractal operators to a generator
$\xi_0({\vec r})$. The structure of the general scattering amplitude is therefore:

\begin{equation}
f_n({\vec q}) = \int d{\vec r} \: e^{i {\vec r} \cdot {\vec q}} \phi_n({\vec r}) ,
\end{equation}

\noindent where quite generally (and similarly for the TD case
with ${\cal F}$ replacing ${\cal G}$),

\begin{equation}
\phi_n({\vec r}) = {\cal A}[{\cal G}^{\dag}_n \xi_0({\vec r})] ,
\end{equation}

\noindent with ${\cal A}$ an operator to be specified next. For
example, in the relatively simple {\em x-ray} case [Eq. (\ref{eq:X})],
${\cal A}$ is the identity; $\xi_0({\vec r}) = n_0({\vec r})$ is the
zero-order local electron density; and $\phi_n({\vec r})$ is the electron
density on the $n^{\rm th}$ iteration of the fractal support. The {\em
electron scattering} case [Eqs.(\ref{eq:Born}),(\ref{eq:e-})] is more
complicated, since now ${\cal A}$ is an integral operator acting on
${\cal G}^{\dag}_n (n_0({\vec r}')/|{\vec r}-{\vec r}'|)$. But the
COVO commutation property, in particular Eq. (\ref{eq:P-int}) with
constant integration limits, ensures that ${\cal G}^{\dag}_n$ can be
taken out and put in front of the integral. The {\em He-scattering}
case [Eqs.(\ref{eq:sud-general}),(\ref{eq:eta})] is probably the most
complicated, since there the fractal operator ${\cal G}^{\dag}$ acts
at several places simultaneously and ${\cal A}$ is an integral
operator with a functional limit.  Nevertheless, the COVO commutation
property and its corollary simplify the problem to the extent that
${\cal G}^{\dag}_n$ may be pulled out again:

\begin{equation}
f_n({\vec Q}) = {1 \over A} \int_A d{\vec R} \: e^{i {\vec R} \cdot {\vec
Q}}\,e^{2i \eta_n({\vec R})} = {1 \over A} \int_A d{\vec R} \: \left( {\cal
G}_n e^{i {\vec Q}\cdot{\vec R}} \right) e^{2i \eta_0({\vec R})} .
\end{equation}

\noindent Indeed, it should now be evident that this is the general
structure of the (local-potential) diffractal-FT problem: the fractal
operator can always be moved from the context-specific part ($\xi_0$) to operate
on the Fourier basis-set, so that generically

\begin{equation}
f_n({\vec q}) = \int d{\vec r} \: \left( {\cal G}_n e^{i {\vec r} \cdot {\vec
q}} \right) \phi_0({\vec r}) .
\label{eq:S-generic}
\end{equation}

\noindent This is the general structure of the Fourier integral: a
context-specific part embodied in the integrand of the form-factor,
$\phi_0({\vec r})$, and {\em a
generic part common to all diffractal-FT problems},
found in the operation of the fractal operator on the Fourier
basis-set. What remains, in order to understand the universal scaling
behavior, is to investigate this latter part.

\subsection{Scaling of the Fourier Integral}
\label{scaling}

The fractal operators to be dealt with here are of the general form given
in the TD and BU cases (Eqs.(\ref{eq:Fd-nD}),Eq. (\ref{eq:Gd}) respectively).
The ensuing analysis closely follows along the lines of the simple, 1D
case treated in Sec.\ref{simple}. 

\subsubsection{TD Case}

Repeating the 1D calculations of Eq. (\ref{eq:apply-F}), one finds that now the application of the fractal operator yields:

\begin{equation}
{\cal F}^n e^{i\,{\vec q}\cdot{\vec r}} = s^{n d} e^{i\,s^n{\vec q}\cdot{\vec
r}} \prod_{j=0}^{n-1} \left( 1+\sum_{i=1}^k e^{i\,s^j\,{\vec q}\cdot{\vec
a}_i\,L} \right) .
\label{eq:F^n-nD}
\end{equation}

\noindent The normalization in the general case is to the relative
volume occupied by the fractal, i.e., $(s^n L)^d (k+1)^n/L^d$, since
there are $(k+1)^n$ fractal elements at the $n^{\rm th}$ iteration,
each with volume $(s^n L)^d$. Introducing a form factor,

\begin{equation}
F({\vec q}) \equiv \int d{\vec r} \: e^{i\,{\vec q}\cdot{\vec r}} \phi_0({\vec
r}),
\end{equation}

\noindent the scattering amplitude assumes the following generic form:

\begin{equation}
f_n({\vec q}) = {1 \over {(s^{n d} (k+1)^n)^{1/2}}} s^{n d} \prod_{j=0}^{n-1}
\left( 1+\sum_{i=1}^k e^{i\,s^j\,{\vec q}\cdot{\vec a}_i\,L} \right)\,
F({\vec q}\,s^n) .
\label{eq:S_n(q)}
\end{equation}

\noindent Thus the exact scaling relation for the intensities reads

\begin{equation}
I_{n+1}({\vec q}) = {s^d \over k+1} \left| 1+\sum_{i=1}^k e^{i\,{\vec
q}\cdot{\vec a}_i\,L} \right|^2\, I_n({\vec q}\,s) ,
\end{equation}

\noindent and by employing the recipe used and justified in
Sec.\ref{simple} for 1D, of evaluating the exponential terms at ${\vec q}=0$,
one obtains the approximate self-affinity relation

\begin{equation}
I_{n+1}({\vec q}) \approx (k+1)s^d \, I_n({\vec q}\,s) .
\end{equation}

\noindent Expressing this through the H\"{o}lder exponent as in the 1D case,
$I_{n+1}(q) \approx s^{-\alpha_1}\,I_n(s\,q)$, one find that the universal
relation between the self-affinity of the intensity spectrum and the fractal
dimension [Eq. (\ref{eq:D-here-nD})], for a TD fractal support, is:

\begin{equation}
\alpha_1 = D-d ,
\label{eq:alpha-2D}
\end{equation}

\noindent in agreement with the 1D case.

\subsubsection{BU Case}

In strict analogy to the results in 1D
[Eqs.(\ref{eq:F^n-BU}),(\ref{eq:S-final-BU})], one finds in the
$d$-dimensional BU case:

\begin{equation}
{\cal G}_n e^{i {\vec q}\cdot{\vec r}} =  s^{-d\,n} {\cal F}^n e^{i s^{-n}
{\vec q}\cdot{\vec r}} = \left[ \prod_{j=1}^{n} \left( 1+ \sum_{i=1}^k e^{i
s^{j-n-1} {\vec q}\cdot{\vec a}_i L } \right) \right] e^{i {\vec q}\cdot{\vec
r}} ,
\label{eq:G^n-dD}
\end{equation}

\begin{equation}
f_n({\vec q}) = {1 \over (k+1)^{n/2}} \left[ \prod_{j=0}^{n-1} \left( 1+
\sum_{i=1}^k e^{i s^{j-n} {\vec q}\cdot{\vec a}_i L } \right) \right]
F({\vec q}) ,
\label{eq:S-final-BU-dD}
\end{equation}

\begin{equation}
F({\vec q}) = \int d{\vec r} \: e^{i {\vec q}\cdot{\vec r}} \phi_0({\vec r}) .
\label{eq:S0-final-BU-dD}
\end{equation}

\noindent The normalization reflects that there are now $(k+1)^n$
elementary units at the $n^{\rm th}$ iteration. Consequently, the scaling of
the intensities is:

\begin{equation}
{I_{n+1}({\vec q}) \over I_0({\vec q})} = {1 \over k+1} \left| 1+\sum_{i=1}^k
e^{i\,s^{-1}{\vec q}\cdot{\vec a}_i\,L} \right|^2\, {I_n({\vec q}/s) \over
I_0({\vec q}/s)} \approx (k+1) \, {I_n({\vec q}/s) \over I_0({\vec q}/s)} =
s^{-\alpha_2} \, {I_n({\vec q}/s) \over I_0({\vec q}/s)} ,
\label{eq:scaling-BU}
\end{equation}

\noindent where again

\begin{equation}
\alpha_2 = D
\label{eq:alpha2=D}
\end{equation}

\noindent with $D$ the fractal dimension [Eq. (\ref{eq:D-here-nD})] of
the BU fractal support. Interestingly, it thus appears that the
embedding space dimension does not enter the scaling in the BU
case. This fact remains to be explained on physical grounds. To
visualize the features of the intensity distribution in this case,
Fig.\ref{fig:self-affine:BU-2D} displays 1D sections of the results of
He scattering calculations from Ag adatoms centered on a sixth
generation Sierpinski carpet with an underlying Pt(111) surface (BU
version of Fig.\ref{fig:2D-fractals}). These results were obtained by
employing Eq. (\ref{eq:S-final-BU-dD}) for the structure factor, and
Eq. (\ref{eq:S0-final-BU-dD}) for the form factor. The latter was
calculated in the Sudden approximation with a realistic He/Ag/Pt
potential \cite{me:Ag-systems} for $\phi_0(\vec{R}) = \exp[2i\,\eta(\vec{R})]$.

To summarize, it was shown that irrespectively of the nature of the probe, for
the scattering of a coherent wave by a deterministic fractal support, the
intensity spectrum is approximately self-affine, with a H\"{o}lder exponent
trivially related to the fractal dimension of the support.

The next generalization, necessary to approach realistic situations,
concerns the effect of randomness.

\section{Scattering From a Randomized Fractal Support}
\label{random}

Realistic fractals always contain some element of randomness
\cite{me:random-model}. For example, in
DLA \cite{Witten} the adsorbing particles perform a random walk and the
resulting fractal is consequently random. Thus it is of major interest to
introduce some randomness into the fractals under consideration, and to
investigate its effect on the conclusions reached so far regarding the scaling
properties of the intensity distribution.  In order to meaningfully introduce
randomness, it is useful to {\em preserve the FD of the support}. Otherwise
the fractal dimension is not a useful descriptor of the scattering
object. This preservation of the fractal dimension can be achieved by keeping
the constant, single contraction factor, but allowing for a {\em distribution
of translations}. The translations will be chosen independently from a given,
but arbitrary, probability distribution $P({\vec a})$, with normalization

\begin{equation}
\int \prod_{i=1}^k d{\vec a}_i^j \: P({\vec a}_i^j) = 1
\label{eq:norm}
\end{equation}

\noindent for each $j$. Here, as before, $j$ is the iteration and $i$ the
translation-number index. The results will of course have to be averaged over
the disorder ensemble, denoted by $\langle \cdots \rangle$ and defined as mean
values over all possible sets $\{ {\vec a}_i^j \}$. Care must be taken to apply
this averaging to the observable {\em intensities} (and not the amplitudes),
since physically, one measures the intensities from a given realization of the
disorder, and averages over the different measurements. Thus:

\begin{equation}
\langle I_n({\vec q}) \rangle = \langle |f_n({\vec q})|^2 \rangle = \int
\prod_{j=0}^{n-1} \prod_{i=1}^k d{\vec a}_i^j \: P({\vec a}_i^j) |f_n({\vec
q})|^2 .
\label{eq:ave}
\end{equation}

In order to visualize the resulting random fractal, it is useful to return
momentarily to the hard-wall, stepped surface language of Sec.\ref{simple}:
The support with randomized translations has steps of constant width as basic
building blocks, but these are spaced randomly over an underlying ``Cantor
grid''. Due to the unequal translations, however, overlaps of steps may now
appear, as illustrated in Fig.\ref{fig:Cantor-step}. It will be shown next
that in the present randomized case, again the intensity spectrum is
self-affine, with the same relations between H\"{o}lder exponent and fractal
dimensions as for the non-random situation.

The change from the deterministic case is that now the fractal operator is
given by

\begin{equation}
{\cal F}^{\dag}_n = \prod_{j=1}^n (\openone + \sum_{i=1}^k {\cal T}_{-{\vec
a}_i^j}\, {\cal C}_{1/s}) ,
\label{eq:Fd-random}
\end{equation}

\noindent where the random shifts $\{{\vec a}_i^j\}$ are chosen from $P({\vec
a})$. Since one still has two identical contractions, the fractal dimension is
unchanged [Eq. (\ref{eq:D-here-nD})], as required.

\subsection{TD Case}

Suppose a measurement is performed on a given random fractal. As for the
calculations leading to the scattering amplitude in the deterministic case
[Eq. (\ref{eq:S_n(q)})], the difference arises in that every translation ${\vec
a}_i$ is replaced by ${\vec a}_i^j$, so that now:

\begin{equation}
f_n({\vec q}) = {1 \over {(s^{n d} (k+1)^n)^{1/2}}} s^{n d} \prod_{j=0}^{n-1}
\left( 1+\sum_{i=1}^k e^{i\,s^j\,{\vec q}\cdot{\vec a}_i^j\,L} \right)\,
F({\vec q}\,s^n) .
\label{eq:S_n(q)-random}
\end{equation}

\noindent The resulting intensities have to be averaged over the disorder
ensemble:

\begin{eqnarray}
\langle I_n({\vec q}) \rangle = {s^{n d} \over {(k+1)^n}} I_0({\vec q}\,s^n)
\Biggl\langle \prod_{j=0}^{n-1} \left| 1+\sum_{i=1}^k e^{i\,s^j {\vec
q}\,{\vec a}_i^j L} \right|^2 \Biggl\rangle  = \nonumber \\
{s^{n d} \over {(k+1)^n}} I_0({\vec q}\,s^n) \prod_{j=0}^{n-1} \int \left[
\prod_{i=1}^k d{\vec a}_i^j\: P({\vec a}_i^j) \right] \left| 1 + \sum_{i=1}^k
e^{i s^j{\vec q}\cdot{\vec a}_i^j L} \right|^2 .
\end{eqnarray}

\noindent For $k=1$ (1D), since the shifts are chosen independently, the last
expression simplifies into a product, and one obtains for the average
intensity:

\begin{equation}
\langle I_n(q) \rangle = s^n I_0(q\,s^n) \,
\prod_{j=0}^{n-1} \left( 1+ \langle \cos(s^j\,q\,a_j L) \rangle \right)
\:\:\: {\rm (1D)}.
\label{eq:I-prod-random-1D}
\end{equation}

\noindent In general, no such simplification occurs, but the scaling
is still tractable:

\begin{equation}
\langle I_{n+1}(q) \rangle = \langle I_n({\vec q} s) \rangle \, {s^d \over
k+1} \int \left[ \prod_{i=1}^k d{\vec a}_i^0\: P({\vec a}_i^0) \right] \left|
1 + \sum_{i=1}^k e^{i {\vec q}\cdot{\vec a}_i^0 L} \right|^2 .
\label{eq:I_n+1-random}
\end{equation}

\noindent In order to express this most accurately in the approximate general
self-affine form of Eq. (\ref{eq:Holder}), the average should be performed at
${\vec q}=0$. Using the normalization condition of the distribution of
translations, Eq. (\ref{eq:norm}) one finds:

\begin{equation}
\langle I_{n+1}(q) \rangle = s^{-\alpha_1}\, \langle I_n(q\,s) \rangle ,
\label{eq:scaling-2-random}
\end{equation}

\noindent with $\alpha_1 = D-d$, just as in the deterministic case
[Eq. (\ref{eq:alpha-2D})] \cite{frac-scatt:comment3}.

\subsection{BU Case}

The scattering amplitude is now given by:

\begin{equation}
f_n({\vec q}) = {1 \over (k+1)^{n/2}} \prod_{j=0}^{n-1} \left( 1+\sum_{i=1}^k
e^{i\,s^{j-n}\,{\vec q}\cdot{\vec a}_i^j\,L} \right)\, F({\vec q}) .
\label{eq:S_n(q)-random-BU}
\end{equation}

\noindent Averaging the intensities over the disorder ensemble:

\begin{equation}
\langle I_n({\vec q}) \rangle = {1 \over (k+1)^n} I_0({\vec q})
\prod_{j=0}^{n-1} \int \left[ \prod_{i=1}^k d{\vec a}_i^j P({\vec a}_i^j)
\right] \left| 1+\sum_{i=1}^k e^{i\,s^{j-n} {\vec q}\,{\vec a}_i^j L}
\right|^2 .
\end{equation}

\noindent The resulting scaling relation is:

\begin{equation}
{{\langle I_{n+1}(q) \rangle} \over {I_0({\vec q})}} = {1 \over k+1} {{\langle
I_n({\vec q}/s) \rangle} \over {I_0({\vec q}/s)}} \, \int \left[ 
\prod_{i=1}^k d{\vec a}_i^n\: P({\vec a}_i^n) \right] \left| 1 +
\sum_{i=1}^k e^{i s^{-1} {\vec q}\cdot{\vec
a}_i^n L} \right|^2 .
\label{eq:I_n+1-random-BU}
\end{equation}

\noindent Performing the average at $q=0$, one obtains:

\begin{equation}
{{\langle I_{n+1}(q) \rangle} \over {I_0({\vec q})}} =
s^{-\alpha_2}\, {{\langle I_n({\vec q}/s) \rangle} \over {I_0({\vec q}/s)}}
\label{eq:scaling-2-random-BU}
\end{equation}

\noindent with $\alpha_2 = D$, again as in the deterministic case
[Eq. (\ref{eq:alpha2=D})].

To conclude, translational randomness alone appears to
have no effect on the scaling properties of the diffraction spectrum.

\section{Further Properties of the Diffraction Spectrum}
\label{form}

The diffraction spectrum is characterized by more than just its scaling
properties. Such features are discussed next.

\subsection{Role of Form Factor}

So far, most of the discussion has centered around the universal scaling
properties of the diffraction spectrum, which were completely determined by
the ``kinematic'' structure factor. However, the role of the ``dynamic'' form
factor cannot be ignored in discussing the properties of the spectrum. It is
in this respect that the different physical probes discussed in
Sec.\ref{methods} differ, and that universality is broken. The form factor
embodies the details of the interaction between probe and scatterer, and
through it the potential enters the intensity spectrum. The example of He
scattering will serve to illustrate the point. In this case, the He/surface
interaction potential enters in a highly non-trivial way
[Eq. (\ref{eq:eta})]. One of the striking consequences is the appearance of
``rainbow'' peaks in the diffraction spectrum \cite{Benny:review2}. These
arise essentially whenever a He atom is scattered from an inflexion point of
the potential (corresponding to maximal force applied to the atom), typically
due to an adsorbed cluster. Following is a brief discussion of the origin and
physical significance of rainbows (see Ref.\cite{me:heptamers} for a more
extensive treatment). It is useful to employ a stationary phase, approximate
evaluation of the Sudden approximation scattering amplitude,
Eq. (\ref{eq:sud-general}). In 1D, the stationary phase condition is:

\begin{equation}
q = -2\,\eta'(x) ,
\label{eq:x(q)}
\end{equation}

\noindent which yields $x(q)$. The scattering amplitude is
then approximated by:

\begin{equation}
f(q) \approx {{e^{i\,q\cdot x(q)} e^{2i\, \eta[x(q)]}} \over {\left| \eta''
\right|_{x(q)}}}
\label{eq:S-classical}
\end{equation}

\noindent The rainbow condition is the existence of an inflexion
point in the phase shift:

\begin{equation}
\eta''(x) = 0 .
\label{eq:eta''}
\end{equation}

\noindent The point $x_0$ satisfying this condition dominates the scattering by
contributing a large peak. In the classical limit of
Eq. (\ref{eq:S-classical}), this shows up as a singularity in the intensity
distribution, at momentum transfer $q_0$ satisfying the stationary phase
condition [Eq. (\ref{eq:x(q)})] together with $x_0$. The singularity of this
crude classical evaluation is smoothed into a finite peak in the more refined
Sudden approximation calculation.

Such Sudden approximation calculations were performed for an Ag/Pt(111) BU
Sierpinski carpet system, with a realistic potential, described in detail in
Ref.\cite{me:Ag-systems}. The results are shown in
Fig.\ref{fig:self-affine:BU-2D} (fractal system) and Fig.\ref{fig:RB} (rainbow
analysis for a single adsorbate). It appears that, although for a single
adatom the rainbow peaks are a dominant feature (Fig.\ref{fig:RB}), in the
case of a fractal system, their role is rather negligible in determining the
structure of the spectrum (Fig.\ref{fig:self-affine:BU-2D}). The reason for
this is that they are far too broad to appear as individual peaks, along with
those due to the fractal support. The rainbows, as well as all other features
of the form factor, act as very broad {\em envelopes} to the detailed spectral
structure. The main effect of the form factor is to provide an overall
intensity decrease, without in any way significantly altering the details of
the structure factor. Since in practice one measures the full intensity
distribution, this can have an effect on its self-affinity properties, and for
a BU fractal care should be taken to divide by the form factor. Similar
results are expected to be found in the diffraction spectra of other probes,
where dynamical factors play an important role, but cannot lead to very peaked
spectral features.

\subsection{``Bragg Conditions'' and Band Structure}

Consider the conditions for maxima derived from the scattering amplitudes for
TD and BU fractals [Eqs.(\ref{eq:S_n(q)}),(\ref{eq:S-final-BU-dD})]. For
TD fractals, the condition is

\begin{equation}
s^j\,{\vec q} \cdot {\vec a}_i = {{2\pi t_i} \over L} ,
\label{eq:Bragg-TD}
\end{equation}

\noindent whereas for BUs, it is

\begin{equation}
s^{j-n}\,{\vec q} \cdot {\vec a}_i = {{2\pi b_i} \over L} .
\label{eq:Bragg-BU}
\end{equation}

\noindent Here $t_i$ and $b_i$ are integers, and $0 \!\leq j \!\leq n-1$. These
are the ``Bragg conditions'' for iteratively generated fractals. However,
since one cannot speak of a conventional unit cell with primitive lattice
vectors in the fractal context, the present conditions for maxima are rather
different from those for periodic crystals. For 1D Cantor-like sets,
Eqs.(\ref{eq:Bragg-TD}),(\ref{eq:Bragg-BU}) reduce to:

\begin{eqnarray}
s^j\,q = {{2\pi t} \over {L a}} \:\:\:\:\: \mbox{TD} \nonumber \\
s^{j-n}\,q = {{2\pi b} \over {L a}} \:\:\:\:\: \mbox{BU}
\label{eq:Cantor-Bragg}
\end{eqnarray}

\noindent Considering first the TD case, the maxima occur for those $q$'s
which, when multiplied by $s^0,s^1,\cdots,s^{n-1}$, are always integer
multiples of ${2\pi \over L\,a}$. For the ternary Cantor set
($s=1/3,\,a=2/3$), with $L=1$, these $q$'s are all the integer multiples of
$3^n \pi$. For $1/s$ equal to an arbitrary {\em integer}, these are the
integer multiples of $(1/s)^{n-1}{2\pi \over L\,a}$. For $1/s$ non-integer,
see Ref.\cite{Allain:2}. Cast in the usual Bragg condition language, $L a
s^{n-1}$ would be an effective ``lattice constant''. The meaning of this
number in the present context, is similar: it is the length of the elementary
building block of the fractal at the $n^{\rm th}$ iteration: the union of
adjacent narrow black and white bars in Fig.\ref{fig:Cantor-step}
(left). However, larger structures also repeat themselves in the fractal, with
smaller frequency. These give rise to the secondary maxima in
Fig.\ref{fig:self-affine}, and mathematically correspond to those $q$'s which
yield integer multiples of ${2\pi \over L\,a}$ for only a subset of
$s^0,s^1,\cdots,s^{n-1}$. The incommensurability of these varying-scale,
repeating structures, is what yields the multitude of peaks in the spectrum,
as opposed to just Bragg peaks in the case of a periodic crystal, and is
ultimately responsible for the self-affinity of the spectrum. The more general
conditions Eqs.(\ref{eq:Bragg-TD}),(\ref{eq:Bragg-BU}), can be interpreted in
a similar fashion.

The distinction between the BU and TD cases is straightforward:
the peak spacings in the former tend to zero (with the peak nearest to the
specular found at ${{2\pi} \over {L a}} s^n$), whereas in the latter the
spacing is unbounded. The only limitation on the position of the furthest {\em
observable} peak in the TD case is energy conservation. In both cases,
however, the structure factors are invariant under a combination of
translations and dilations (apart from the reduction in intensities,
responsible for the self-affine properties). One is thus led to define a new
basis of primitive vectors for the reciprocal space, from which a Brillouin
zone can be constructed. As seen in Fig.\ref{fig:self-affine}, the regions
connected by these operations do not overlap, and can be considered as
separate {\em bands}. A detailed treatment of this point is given in
Ref.\cite{Allain:2}, and will not be repeated here.

\section{Self-Affine or Power-Law?}
\label{SA-power}

As mentioned in the Introduction, the common wisdom relating to scattering by
{\em random} fractal objects (e.g. porous solids \cite{Schmidt:1,Wong:1}), amply confirmed experimentally, is
that close to the specular the intensity satisfies a power law:

\begin{equation}
I(q) \approx q^{-\gamma}
\label{eq:power}
\end{equation}

\noindent with $\gamma = D +$const. This power-law decay is clearly very
different from the self-affine intensity spectrum predicted here for iterative
fractals. Considering the unquestionable experimental evidence for the
power-law, this discrepancy calls for clarification. The following
arguments may shed some light on this issue.

In order to derive the power-law [Eq. (\ref{eq:power})], one typically starts with the
definition of a ``mass fractal dimension'', describing the scaling of the mass
$N(r)$ enclosed in a sphere of radius $r$, centered at an arbitrary point in
the fractal:

\begin{equation}
N(r) \approx r^D .
\label{eq:N(r)}
\end{equation}

\noindent If the fractal is self-averaging (an assumption which is implicit in
the derivation of, e.g., Refs.\cite{Schmidt:1,Wong:2}), then this mass is
related to the pair distribution function $g(r)$ by

\begin{equation}
N(r) = \langle \rho \rangle \int_0^r g(r') d^dr' ,
\label{eq:N-g}
\end{equation}

\noindent with $d$ the embedding space dimension and $\langle \rho \rangle$
the average density. From general scattering theory it is known that the
structure factor is

\begin{equation}
S({\vec q}) = 1+\langle \rho \rangle \int [ g(r)-1 ] e^{i {\vec q} \cdot {\vec
r} } d^dr .
\label{eq:S(q)}
\end{equation}

\noindent From here one arrives at Eq. (\ref{eq:power}) (see Ref.\cite{Teixeira}
for details).

It is thus seen that the crucial assumption invoked in this derivation is the scaling law
Eq. (\ref{eq:N(r)}). It must be realized, however, that this expression is in
many cases only true {\em on average}. This can be seen very clearly for the
ternary Cantor set (Fig.\ref{fig:Cantor-step}). Suppose the set has bars of
unit height and one calculates its
cumulative mass $N^{(j)}$ in the $j^{\rm th}$ iteration, starting from the
left, and in terms of the number of black bars. Then the following recursion
formula may easily be verified:

\begin{equation}
N^{(j+1)} = N^{(j)} \cup \{ \left| N^{(j)} \right| {\rm times} [{\rm
last}(N^{(j)})] \} \cup \{ N^{(j)}_l + [{\rm last}(N^{(j)})]
\}_{l=1}^{\left|N^{(j)}\right|} \:\:\:\:\: N^{(0)} = \{1\} .
\label{eq:recursion}
\end{equation}

\noindent Here $\left|N^{(j)}\right|$ is the length of the sequence $N^{(j)}$
and last$(N^{(j)})$ is its last term. Indeed, the zeroth-iteration ternary
Cantor set consists of 1 black bar, the first iteration has a cumulative mass
of $\{1,1,2\}$ black bars, the second iteration has mass
$\{1,1,2,2,2,2,3,3,4\}$, etc. In Fig.(\ref{fig:SA-power}), $N^{(7)}$ is
displayed on a log-log plot, together with the power-law $N(r)$ [as suggested
from Eq. (\ref{eq:power})], i.e., a line with slope ln2/ln3, the fractal
dimension of the ternary Cantor set. It can be seen that this line serves as
an accurate {\em envelope} to the actual $N(r)$, which is in fact a ``Devil's
staircase'', with a very rich (fractal) structure. This example illustrates
the general situation: A simple scaling law of the form of
Eq. (\ref{eq:power}) is only an {\em average} representation of the actual
cumulative mass function of a fractal, which may in fact not be self-averaging. Since in this work the exact
properties of the fractal [i.e., equivalent to Eq. (\ref{eq:recursion})] were
used to calculate the scattering intensities, it should now come as no
surprise that the resulting diffraction spectra themselves displayed the full,
rich structure of the scattering fractal object. Conversely, had the power-law
form of Eq. (\ref{eq:power}) been used in the present scattering calculations,
the result would have been a power-law decay of the intensity.

Why then do experiments from natural fractals yield the power-law? The
preceding arguments strongly suggest that this is related to an averaging
process which smoothes the fine-structure of the intensity distribution. A
priori, two types of averages could be considered: (1) over the position of
the center point of the cumulative-mass calculation, and (2) over the disorder
ensemble. The first type can be ruled out immediately, however, since it is
common to both the power-law and self-affine spectra derivations: The
calculation of an {\em intensity} involves a double integral in which {\em
all} pairs of points appear in the form $\int\int dr\,dr'\: \exp[i\,q(r'-r)]
n(r)\,n(r')$. This automatically performs the first type of average. Thus, by
elimination, the ensemble average appears to be responsible for smoothing out
the self-affine properties into a simple power-law decay. Indeed, in the
deterministic iterative fractals considered here, there is of course no
ensemble to average over, in contrast to the typical experimental situation.
Interestingly, the random fractals of Sec.\ref{random} are ``not random
enough'', since they also display a self-affine spectrum. The type of
randomness encountered in experiments yielding the power-law must lead, in
contrast, to self-averaging between the physical cut-offs. There is a further
difference between the randomness considered here and that encountered in
experiments, namely that in the latter the randomness does not preserve the FD
above the upper cut-off. In contrast, the type of randomness considered in
this work preserves the FD on all scales. It is possible that this difference
plays a role in creating the discrepancy between the experimental power-law
results and the theory presented here.

\section{Conclusions}
\label{conclusions}

In summary, the diffractal-Fourier transform problem, for scattering of
coherent waves from a wide class of iteratively constructed fractals, was
solved analytically, yielding the scaling properties of the diffraction
spectrum. The class of fractals considered here is not that which is typically
observed in scattering experiments, and is characterized by a self-affine
intensity spectrum. A simple relation was found to exist between the
self-affinity exponent of this spectrum and the fractal dimension of the
scattering fractal support. In contrast, many experiments yield intensity
distributions characterized by a power-law decay. It is argued here that this
is predominantly the result of scattering from {\em self-averaging} random
fractals, which are more abundant in experimental realizations of
fractality. The results apply to a large variety of scattering probes, from
neutron to He scattering, the condition being the applicability of the Fourier
transform. The differences among the probes are contained in a form factor,
which, however, does not seem to have an important role in determining the
details of the diffraction spectrum. It would be of interest to see whether
scattering from non-self-averaging (random) fractal systems will yield a
self-affine intensity spectrum as predicted here. Further theoretical work
will concentrate on generalizing the types of randomness studied here, and on
investigating the possible role of cut-offs in leading to the power-law decay
of the intensity observed in many experiments.

\acknowledgements

I would like to acknowledge most helpful discussions with Dr.
Leonid Baranov, stimulating comments by Profs. Ofer Biham and R. Benny Gerber,
and permission to use unpublished He scattering calculations from Dr. Tamar
Yinnon.

\section*{FIGURE CAPTIONS}
\begin{figure}
\caption{Left: Third and fourth iterations of a step on a ternary Cantor
set support. Middle and right: Same, but with randomized translations.}
\label{fig:Cantor-step}
\end{figure}

\begin{figure}
\caption{Generator and first two iterations of the top-down Sierpinski carpet,
supporting a harmonic potential well (contour lines). By expanding each
iteration so that every square is of unit size, the corresponding bottom-up
fractal can be obtained.}
\label{fig:2D-fractals}
\end{figure}

\begin{figure}
\caption{Test of the scaling relation for TD fractals
[Eq. (\protect\ref{eq:scaling-2})]: Superimposed intensities (arbitrary units),
from Eq. (\protect\ref{eq:I_n(q)-TD}), for He scattering by a hard-wall step
function, built on the fifth and sixth iterations of a ternary (top) and
generalized {\em TD} Cantor set with $s=1/5$, $a=4/7$ (bottom). The
intensity from the fifth iteration (dotted line) is rescaled according to
Eq. (\protect\ref{eq:scaling-2}). Clearly, the rescaled intensity serves as an
accurate envelope. In the $n \protect\rightarrow \protect\infty$ limit,
therefore, subsequent iterations become indistinguishable and the intensity is
self-affine. The insets show magnifications, in which a coarse-grained
reproduction of the entire peak structure can be identified, illustrating the
self-similarity of the spectrum.}
\label{fig:self-affine}
\end{figure}

\begin{figure}
\caption{Test of the scaling relation for BU fractals
[Eq. (\protect\ref{eq:scaling-2-BU})]: Superimposed structure factors
(arbitrary units), from Eq. (\protect\ref{eq:S-final-BU}), for He scattering by
a hard-wall step function, built on the same Cantor sets as in
Fig.\protect\ref{fig:self-affine}. The intensity from the fifth iteration
(dashed line) is rescaled according to Eq. (\protect\ref{eq:scaling-2-BU}).
Again, the rescaled intensity serves as an accurate envelope, although the
agreement worsens with increasing $q$. Insets as in
Fig.\protect\ref{fig:self-affine}.}
\label{fig:self-affine:BU}
\end{figure}

\begin{figure}
\caption{Results of epsilon-variation analysis \protect\cite{Dubuc} of the
intensities displayed in Fig.\protect\ref{fig:self-affine:BU}. The slope of
the log-log plots yields the self-affinity exponent as 0.633 for the $n=8$
ternary Cantor set and 0.41 for the $n=6, s=1/5$, $a=4/7$ set.}
\label{fig:variation}
\end{figure}

\begin{figure}
\caption{Top: Structure factor for He scattering from Ag adatoms on a Pt(111)
surface, with the Ag adatoms positioned on BU, sixth iteration
Sierpinski carpet. The generator is a Pt(111) unit-cell ($L=2.77\AA$).  The
self-similar structure can be noticed upon careful examination.  Bottom: The
complete intensity spectrum, after multiplication by the form factor
(Fig.\protect\ref{fig:RB}). The effect is mainly an overall intensity decrease
with increasing $q$. The rainbows are too broad to be noticed as individual
peaks.}
\label{fig:self-affine:BU-2D}
\end{figure}

\begin{figure}
\caption{Top: Classical turning points for a single Ag atom adsorbed on a flat
Pt(111) surface, for He at normal incidence with
$k_z=6\AA^\protect{-1\protect}$. The inflexion points are indicated (1-3),
along with rays (guide for the eye only), indicating the trajectories of
classical particles scattered from these points (note the difference in scale
between the axes, causing the apparently non mirror-like reflection). In the
hard-wall approximation [Eq. (\protect\ref{eq:hw})], the inflexion points
coincide with those of the phase-shift function, and approximately yield the
positions of the rainbow peaks through the stationary phase and singularity
conditions [Eqs.(\protect\ref{eq:x(q)}),(\protect\ref{eq:eta''})]. Using this,
the scattering angles are found to be (1) $21.3^\protect{\circ\protect}$ and
(2) $2.4^\protect{\circ\protect}$ with respect to the normal to the surface,
corresponding to $q=2.2 \AA^\protect{-1\protect}$ and $0.25
\AA^\protect{-1\protect}$, which are approximately the rainbow positions
indicated in the scattering intensity (bottom).}
\label{fig:RB}
\end{figure}

\begin{figure}
\caption{Log-log plots of the exact cumulative-mass relation
[Eq. (\protect\ref{eq:recursion})] for a seventh
iteration ternary Cantor set, and a power law with exponent equal to this
set's fractal dimension [Eq. (\protect\ref{eq:power})].}
\label{fig:SA-power}
\end{figure}


\begin{thebibliography}{10}
\bibitem{frac-scatt:comment0}
{It should be noted that these are {\em not} models of self-affine, rough
  surfaces, also considered in the literature \cite{Wong:2}.}
\bibitem{frac-scatt:comment4}
{An example of a non-local potential is the so-called separable, or factorable
  potential $V=|\zeta\rangle \langle\zeta |$, where $\zeta$ is some fixed
  vector in the one-particle Hilbert space. Non-locality appears through
  $\langle \vec{x}'|V|\vec{x} \rangle = \zeta(\vec{x}')\zeta(\vec{x}^{*})$,
  unlike $\delta(\vec{x}'-\vec{x})$ for the matrix element of local
  potentials.}
\bibitem{frac-scatt:comment5}
{A much richer class of dilations is possible by employing a general linear
  transformation, i.e., representing $s$ by matrix. Not all such
  transformations can be conveniently represented in exponential form with a
  single exponential, but many of them can. In particular, $\exp(x_i S_{ij}
  \partial_{x_j}) f(x) = f(s x), s=\exp(S)$. This general $s$ is capable of
  providing a simultaneous rotation and stretch along arbitrarily directed axes
  with individual factors for each axis.}
\bibitem{frac-scatt:comment1}
{A similar result was derived by Allain and Cloitre \cite{Allain:2,Allain:1},
  in the case of optical scattering in the eikonal approximation, for
  doubly-infinite BU fractals.}
\bibitem{frac-scatt:comment3}
{It is tempting to consider also the polydisperse case, where dilations are
  chosen at random from a given probability distribution $P(s)$ (which is far
  more difficult). In this case one deals with a fractal operator of the type
  ${\cal F}^{\dag}_n = \prod_{j=1}^n (\openone + {\cal T}_{-a}\, {\cal
  C}_{1/s_j})$, for which it is still possible to obtain an explicit expression
  for the scattering intensities, using the ideas of the preceding
  calculations. This will, however, not be pursued any further here, since for
  random dilations the fractal dimension can only be defined {\em on average},
  and thus ceases to be a useful characteristic of the scattering set: From
  Eq. (\ref{eq:D-here}), $\langle D \rangle = \ln(2) \int ds\: P(s)
  1/\ln(1/s)$.}
\end{thebibliography}
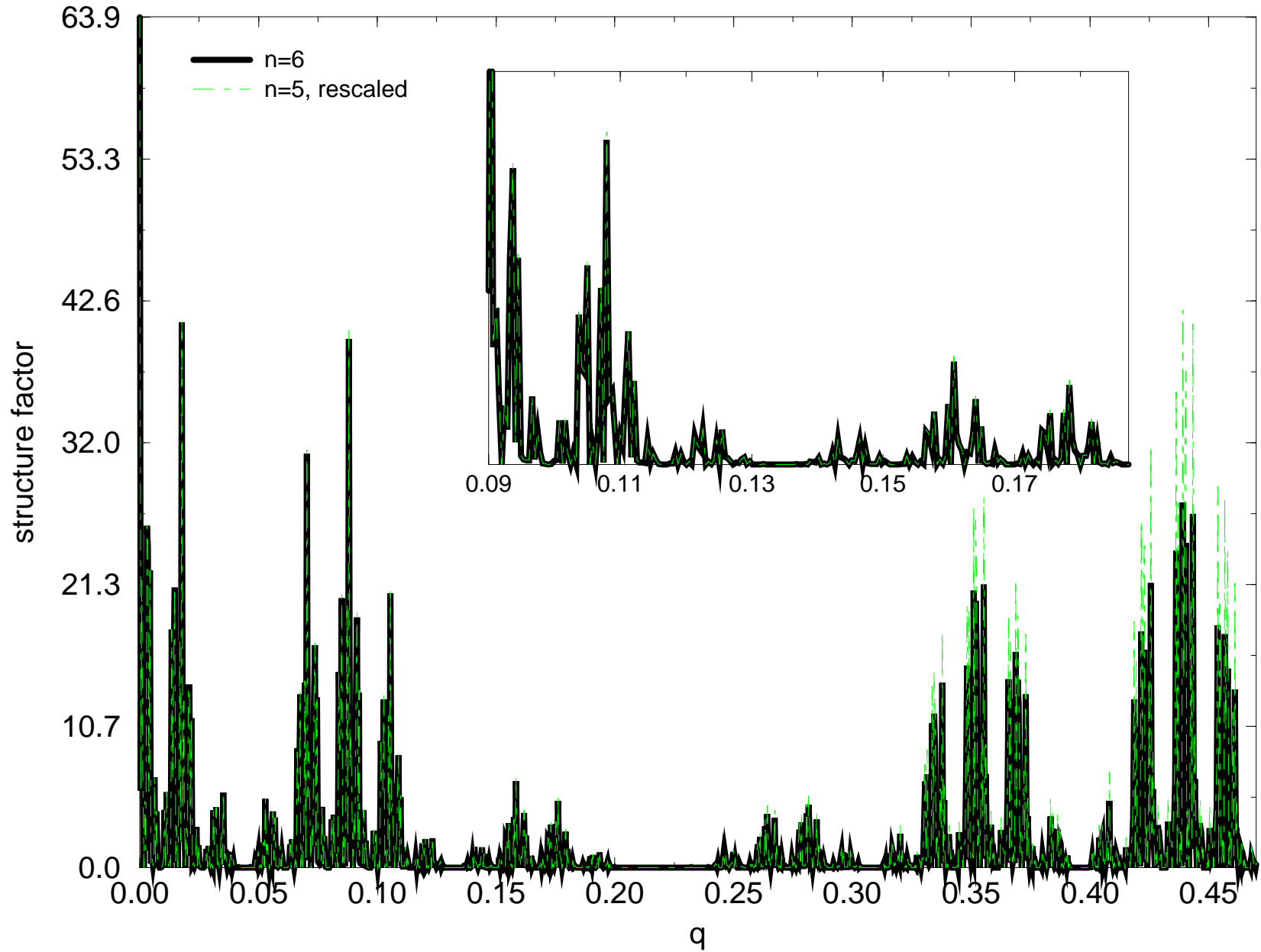
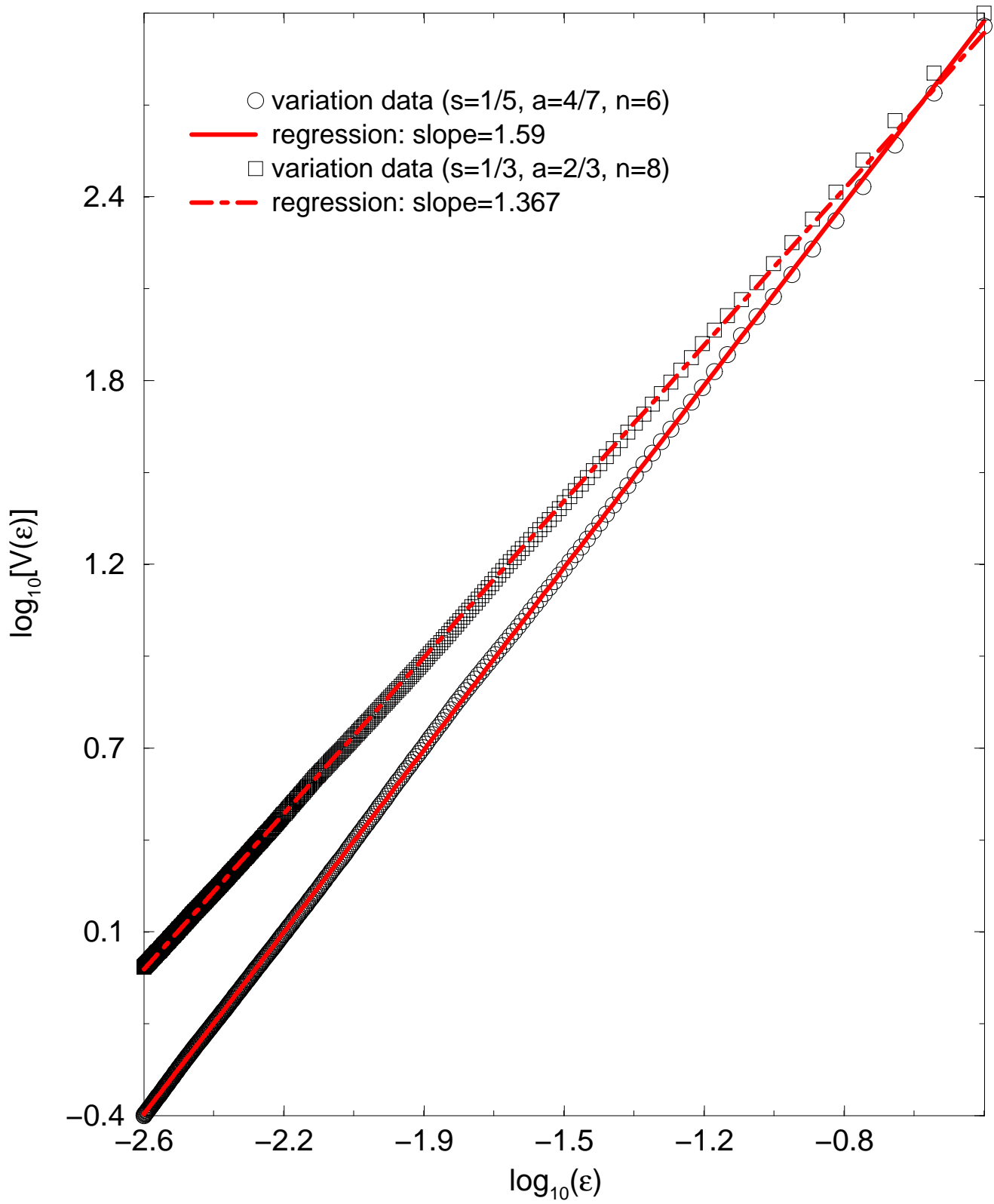
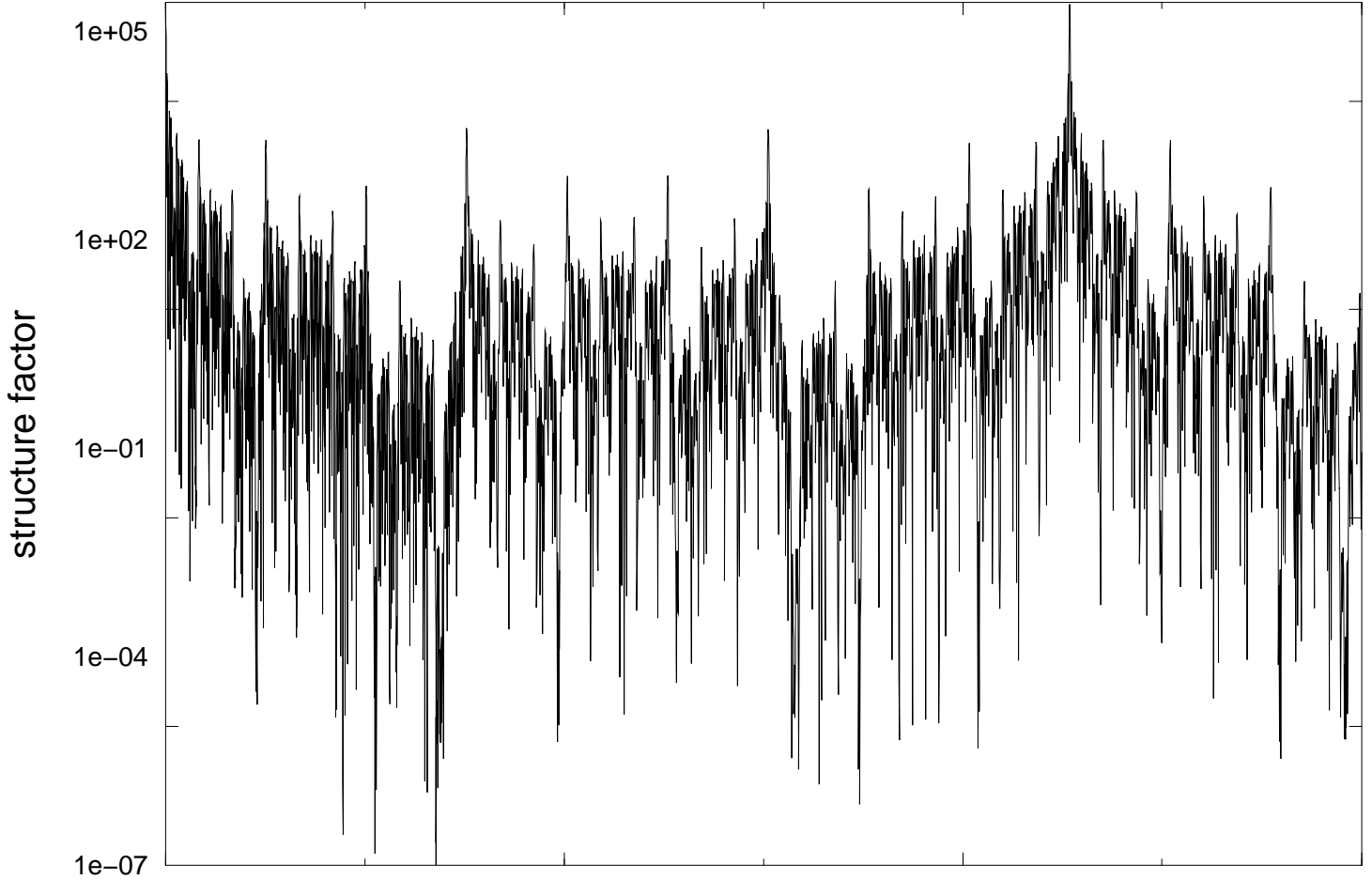
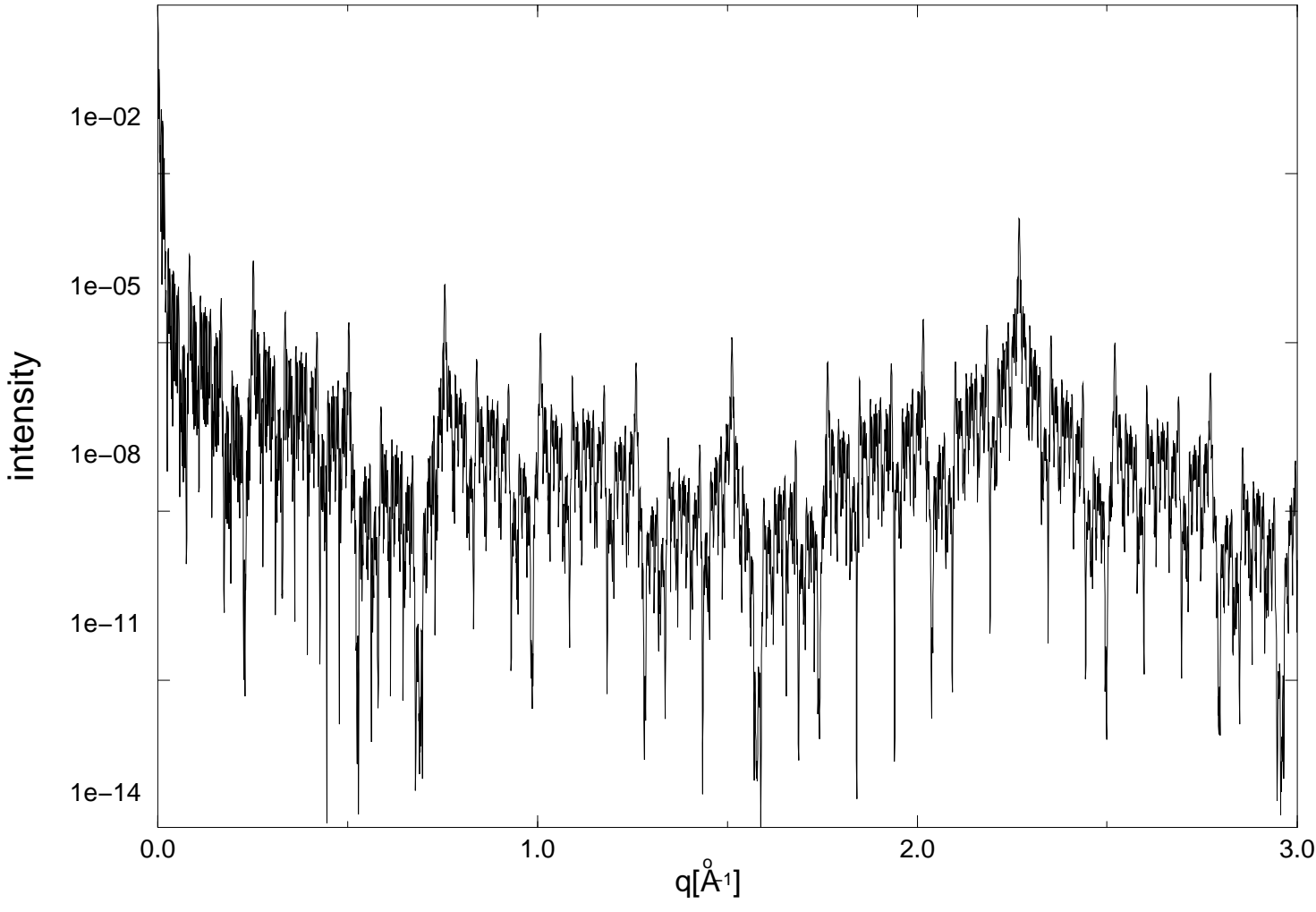
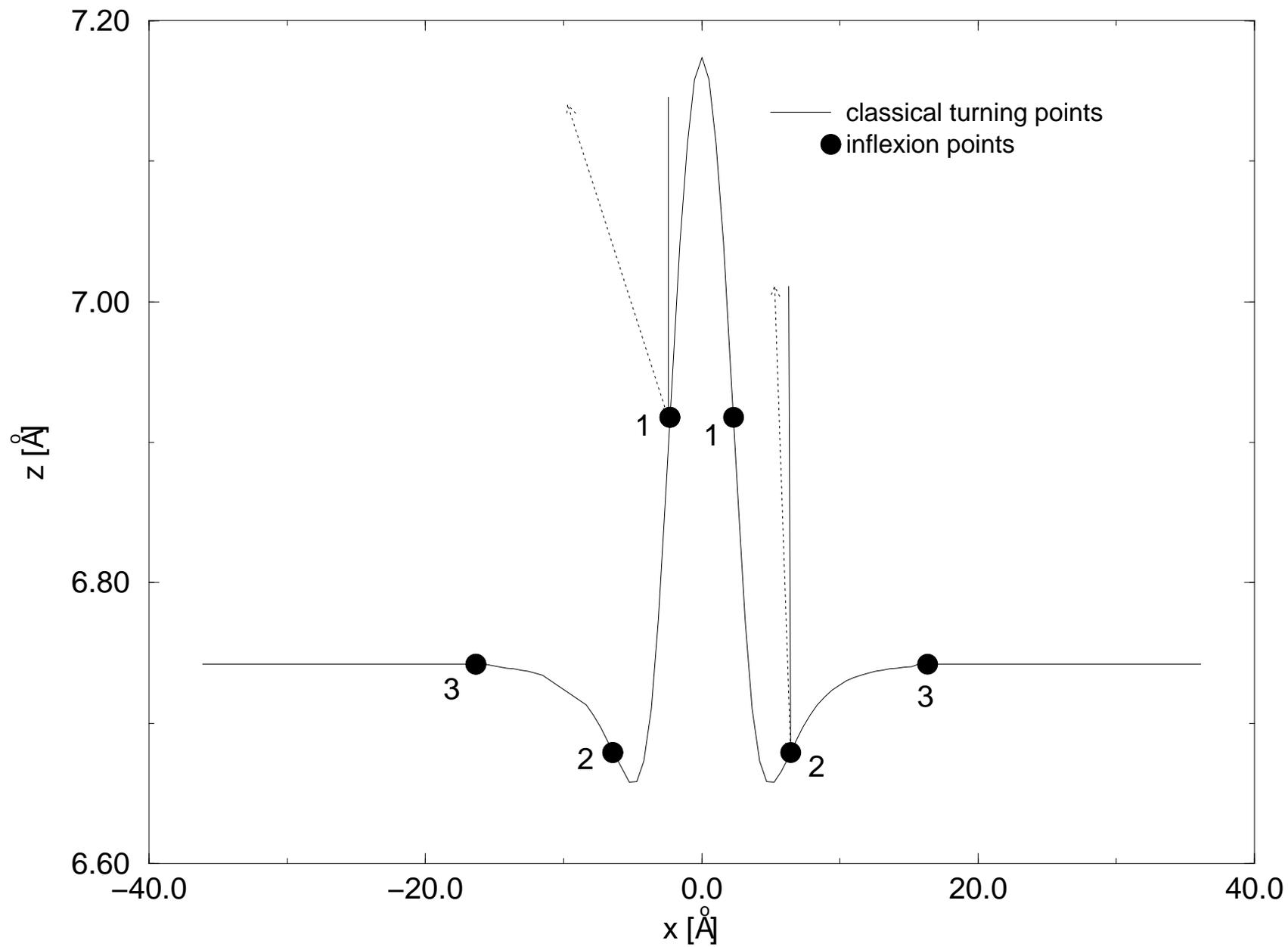
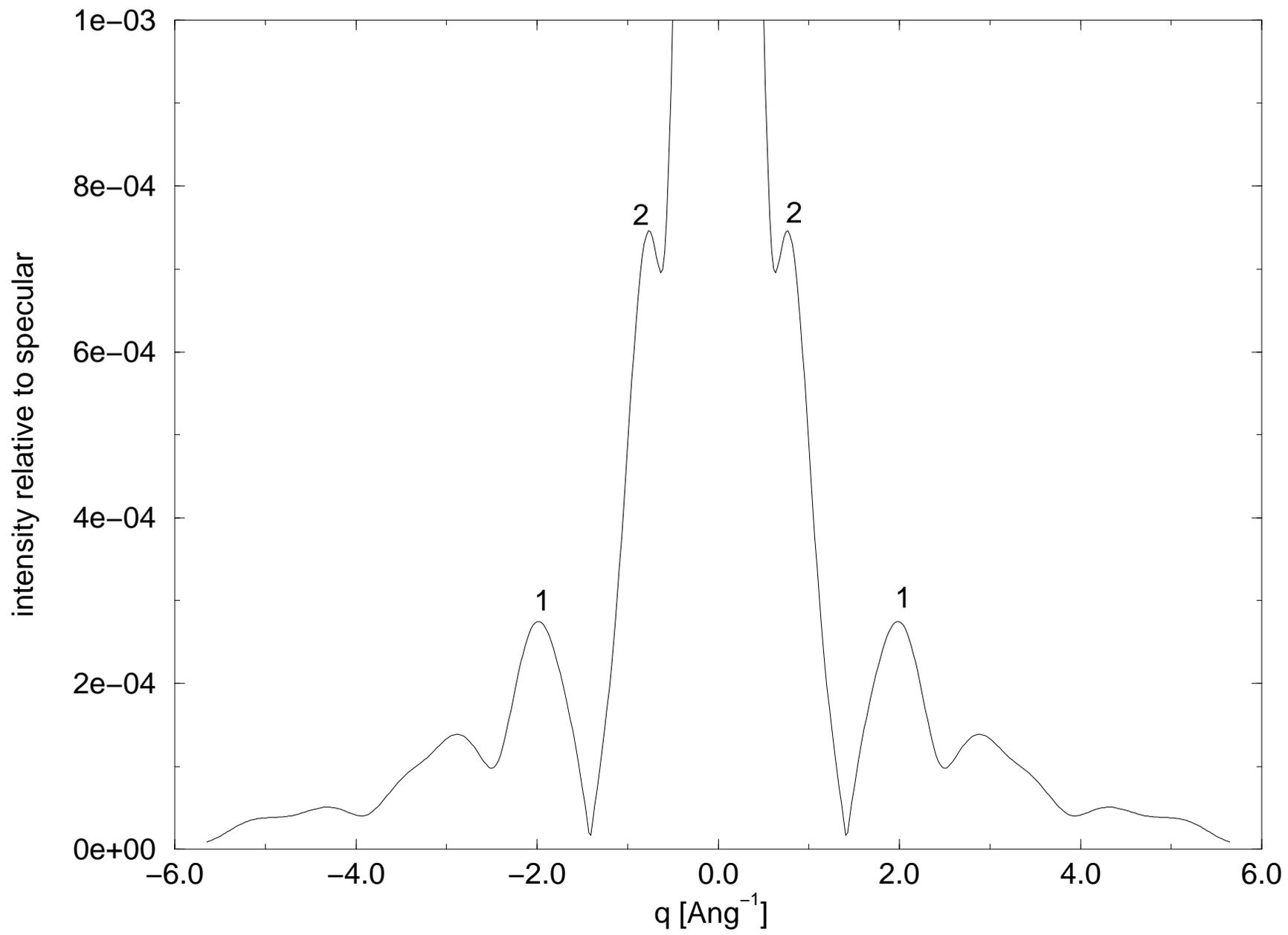
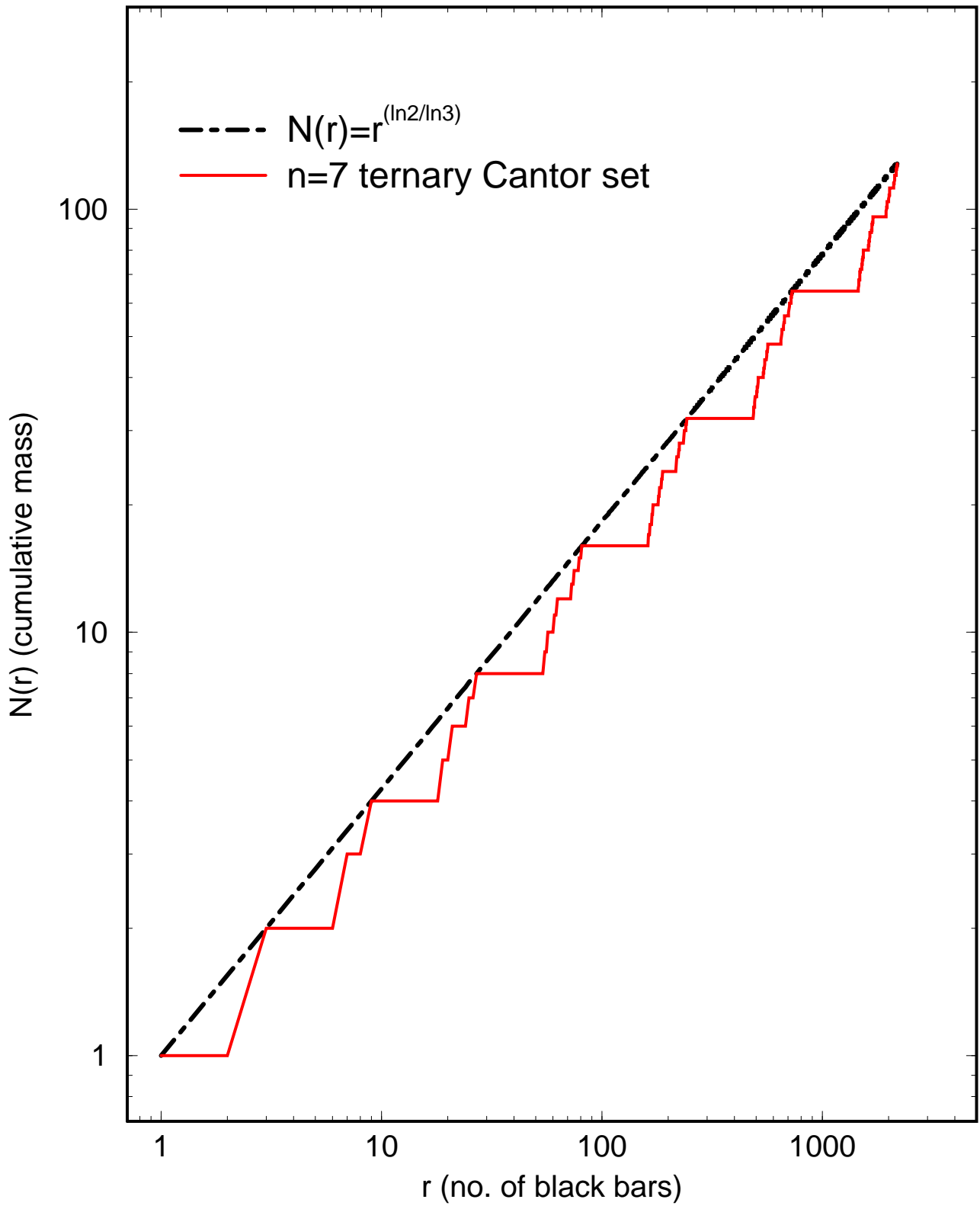
\end{document}